\documentclass[%
 reprint,
superscriptaddress,
 amsmath,amssymb,
 aps,
]{revtex4-2}

\usepackage{graphicx}
\usepackage{dcolumn}
\usepackage{bm}
\usepackage{siunitx}
\usepackage{pstricks}
\usepackage{pgf}
\usepackage{caption}
\usepackage{comment}
\usepackage{booktabs}
\usepackage{setspace}

\begin{document}

\preprint{APS/123-QED}

\title{Isotope shift measurement of the $423$-$\SI{}{\nano\meter}$ transition in neutral Ca}

\author{David Röser}
 \email{roeser@physik.uni-bonn.de}
 \affiliation{Physikalisches Institut, Universität Bonn, Nussallee 12, 53115 Bonn, Germany}
\author{Lukas Möller}%
\affiliation{Physikalisches Institut, Universität Bonn, Nussallee 12, 53115 Bonn, Germany}
\author{Hans Keßler}
\affiliation{Physikalisches Institut, Universität Bonn, Nussallee 12, 53115 Bonn, Germany}
\author{Simon Stellmer}
 \email{stellmer@uni-bonn.de}
\affiliation{Physikalisches Institut, Universität Bonn, Nussallee 12, 53115 Bonn, Germany}

\date{\today}

\begin{abstract}
We report on saturated absorption spectroscopy measurements of the $(4s^2) ^{1}S_0\rightarrow(4s4p) ^{1}P_1$ transition for the four most abundant even-mass isotopes in calcium. By referencing the laser locked to an ultralow expansion cavity and carefully investigating systematic errors, isotope shifts are determined with a precision below $\SI{100}{\kilo\hertz}$, improving previously reported values by a factor of about five. A King plot analysis employing literature values of the $729$-$\SI{}{\nano\meter}$ transition in $\mathrm{Ca}^+$ ions shows excellent linearity. The field and mass shift parameters are determined from King plots with other transitions.
\end{abstract}

\maketitle

\section{\label{sec:intro} Introduction}

Laser spectroscopy plays a key role in investigating the atomic or molecular structure \cite{demtröder2015laser2}. Over the past decades, spectroscopic techniques have been developed and improved to outstanding precision on the single $\SI{}{\hertz}$ level \cite{Ludlow2015,Knollmann2019, Micke2020, Hur2022, Ono2022, Chang2024}. These measurements can be used to define frequency standards or explore physics beyond the standard model, \textit{e.g.}, to search for temporal variations in fundamental constants \cite{Berengut2011}, breaking of local Lorentz invariance \cite{Dreissen2022,Ohayon2022}, CPT violation \cite{Ohayon2022}, or a new gauge boson \cite{Delaunay2017,Solaro2020,Counts2020,Berengut2020,Rehbehn2021,Hur2022,Figueroa2022,Potvliege2023}.

The term isotope shift refers to the energy difference of electronic states between different isotopes. The main contributions to the isotope shifts are the mass and the field shift. The mass shift originates from a change of the electron's reduced mass and momentum correlations of electrons and is proportional to the relative mass difference between isotopes. The field shift arises from a different nuclear charge distribution due to different neutron numbers \cite{Flambaum2017}.

In a simplified case, the field shift is assumed to be proportional to the difference of the squared nuclear charge radii. Then, isotope shifts of different transitions in atoms or ions of the same species are linearly related to each other, when they undergo a King plot analysis \cite{King1963}. Deviations from the linear behavior allow insights into the nuclear potential, as higher-order corrections or new physics terms might be determined \cite{Berengut2018,Hur2022}.

A possible new boson might be a force carrier for an interaction between neutrons and electrons. Their coupling would change the nuclear potential by a Yukawa-type potential, which is dependent on the neutron number \cite{Counts2020}. Hence, the interaction could be visible in isotope shift measurements and appear as a nonlinearity in a King plot \cite{Berengut2018,Rehbehn2021,Flambaum2017}.

It has been found that strong nonlinearities can occur due to large nuclear deformation. Here, the simple field shift approximation fails, as the quadratic field shift and the next higher order Seltzer moment have to be taken into account \cite{Counts2020,Hur2022}. Still, the combination of various transitions in the Yb system allowed to constrain the parameter space for the new boson coupling \cite{Ono2022}. An overview of the work on $\mathrm{Yb}$ can be found in \cite{Door2024}.

For isotope shift measurements, group II elements are of particular interest. Their even-mass isotopes feature an even number of protons and an even number of neutron leading to zero nuclear spin. The absence of hyperfine structure circumvents center-of-gravity determinations and thus enhances measurement precision.

Among the group II elements, calcium holds particular significance, as it offers two 'doubly magic' nuclei. They feature a closed proton and a closed neutron shell, resulting in non-deformed nuclei \cite{Nakada2019}. Isotope shift measurements on neutral or ionized Ca have been measured on numerous transitions with different methods, \textit{e.g.}, \cite{Knollmann2019,Solaro2020, Mortensen2004,Nörtershäuser1998,Andl1982,Salumbides2011,Rehbehn2021,Gebert2015,Micke2020,Chang2024}. Especially the shifts on the $4s^2S_{1/2}\rightarrow 3d {}^{2}D_{5/2}$ quadrupole transition at $\SI{729}{\nano\meter}$ in $\mathrm{Ca}^+$ have been measured with outstanding precision \cite{Knollmann2019,Chang2024}. Reported King plots show excellent linearities, further narrowing the bounds on the coupling strength of a new possible boson \cite{Chang2024}. An overview about previous work up to 2020 can be found in Ref.~\cite{Kramida2020}.

The $(4s^2) {}^{1}S_0\rightarrow(4s4p) {}^{1}P_1$ transition in Ca I is at $\SI{423}{\nano\meter}$ with a natural linewidth of $\SI{34.5(4)}{\mega\hertz}$ and saturation intensity of $\SI{54(5)}{\milli\watt\per\centi \meter^2}$. Isotope shifts on this line have been measured with different methods, \textit{e.g.}, ionization spectroscopy \cite{Nörtershäuser1998} and fluorescence detection of an atomic beam \cite{Andl1982,Salumbides2011}. The values reported have typical $1\sigma$ uncertainties of several $\SI{100}{\kilo\hertz}$. While the reported shifts for $^{44}\mathrm{Ca}$ and $^{48}\mathrm{Ca}$ agree very well, there is a $\SI{400}{\kilo\hertz}$ discrepancy for $^{42}\mathrm{Ca}$ \cite{Nörtershäuser1998,Salumbides2011}.

In this article, we report on a saturated absorption spectroscopy measurement in a gas cell, narrowing the uncertainty on the isotope shifts to the sub-$\SI{100}{\kilo\hertz}$ range. A King plot analysis proves linearity between our measurement and isotope shifts on the $729$-$\SI{}{\nano\meter}$ line in $\mathrm{Ca}^+$ \cite{Knollmann2019}. The field and mass shift coefficients for the $423$-$\SI{}{\nano\meter}$ lines are determined from isotope shift data on dipole transitions in $\mathrm{Ca}^+$ \cite{Gebert2015}.

\section{\label{sec:satabs} Experimental Apparatus}

\subsection{Optical setup}
The laser light used in the experiment is derived from an infrared diode laser system equipped a with tapered amplifier operating at $\SI{846}{\nano\meter}$ (Toptica TA pro). A vibration isolated and temperature stabilized ultra-low expansion (ULE) cavity (Menlo Systems ORC-Cubic) serves as a frequency reference. The laser undergoes a carrier shift modulation and the sideband is frequency-stabilized with the Pound-Drever-Hall method to the ULE cavity. The carrier shift frequency can be tuned with a direct digital synthesis generator with a $\SI{}{\hertz}$ level precision. The performance of the lock has been evaluated by locking two identical lasers with different carrier shifts to the same cavity mode. Their beat signal linewidth is smaller than $\SI{2}{\hertz}$ on a timescale of $1$ second.


The light is converted to $\SI{423}{\nano\meter}$ with second harmonic generation (SHG) in a nonlinear crystal within an optical cavity, which is locked to the laser using the Hänsch-Couillaud method \cite{Hansch1980}. A polarization maintaining fiber is used to clean the mode profile and guide the linearly polarized light to the experiment table.

To suppress intensity noise in the experiment, the light power is measured on the experiment table. A PID controller is used to stabilize it by applying feedback on the tapered amplifier current. The locking bandwidth is measured to approximately $\SI{80}{\kilo\hertz}$, which is mainly limited by the slow response of the SHG cavity.

The laser light is split into two beams of equal intensities ($\SI{1.9}{\milli\watt}$ on a Gaussian beam diameter of $\SI{4}{\milli\meter} $), with one serving as pump and the other as the probe beam. Saturated absorption spectroscopy is performed by sending both beams counter-propagating through a $\SI{1.5}{\meter}$ long stainless steel vacuum system acting as a gas cell \cite{demtröder2015laser2}. Lenses are slightly tilted from normal incidence and all other optics, including the vacuum viewports, are wedged to avoid ethaloning effects due to reflections or stray light impinging on the photodiodes or interacting with the probed velocity class of atoms. 

A lock-in technique is employed by sinusoidal amplitude modulation of the pump beam with an acousto-optic modulator (AOM). The lock-in frequency is set to $\SI{11}{\kilo\hertz}$, as the the spectral noise amplitude is minimal here and all electronic components reach the white noise floor.

The AOM is driven at a frequency of $\SI{200}{\mega\hertz}$. Thus, interference effects between pump and probe beam are modulated much faster than the atomic linewidth. To reduce residual Doppler shifts, proper overlap of pump and probe beam is ensured by coupling the probe beam into the AOM and observing the diffraction orders \cite{Park2001}.

\begin{figure}
\includegraphics[width=0.47\textwidth]{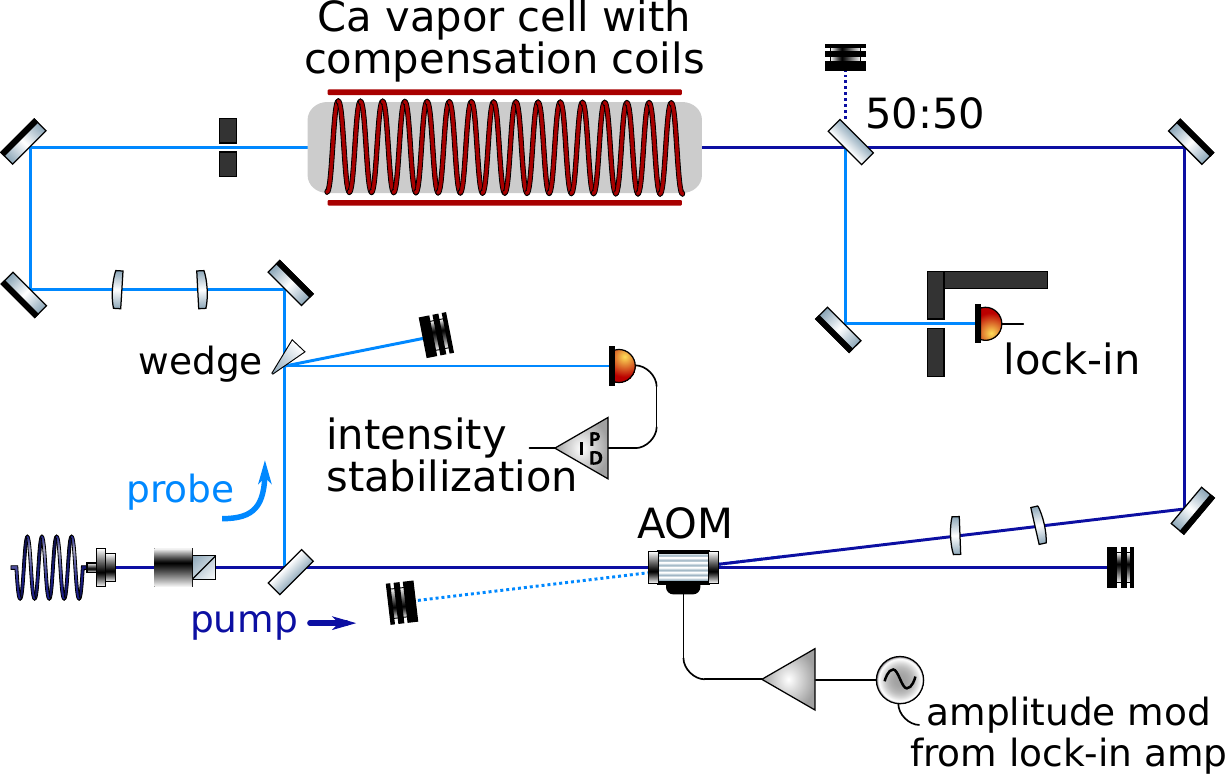}
\caption{Sketch of our saturated absorption spectroscopy setup.}
\label{fig:spec}
\end{figure}

The Lamb dip is measured with a photodiode followed by a low noise transimpedance amplifier (FEMTO DHPCA-100), which is then connected to a lock-in amplifier (Zurich Instruments HF2LI). The demodulated signal is low pass filtered, typically with a time constant of $5$ seconds. Measurements are performed by shifting the laser frequency and a subsequent settling time of twice this time constant.  Afterwards, the lock-in signal is recorded and used for evaluation. This procedure leads to a gain in signal-to-noise ratio (SNR) of four orders of magnitude compared to unmodulated spectroscopy. A sketch of the optical setup is shown in Fig.~\ref{fig:spec}.

\subsection{Systematic effects}

Five grams of naturally abundant calcium are placed into a $\SI{1.1}{\meter}$ long reservoir region, which is kept at $\SI{340}{\degreeCelsius}$ by a magnetic field compensating heating tape. The temperature inhomogeneity is monitored with sensors and measured to be smaller than $\SI{10}{\degreeCelsius}$.

The vapor pressure of Ca is estimated to be $\SI{e-7}{\milli\bar}$, leading to approximately $\SI{20}{\percent}$ absorption of each beam \cite{Alcock1984}. At this level, Beer-Lambert's law can be utilized to show that the combined light power of probe and pump beam remains constant within the uncertainty margin of our power measurement. This allows the determination for AC Stark shifts: Because of the Doppler broadened absorption in the gas cell, the different isotopes are measured with different power and thus have to be corrected. 

To further minimize systematic errors, any magnetic fields including the Earth's magnetic field are compensated with coils in all directions. The remaining field is significantly smaller than $\SI{5}{\mu\tesla}$ and reduces Zeeman splitting and thus polarization dependent effects.

In test measurements, pump and probe beam have been deliberately set to a significant power imbalance to enhance light pressure-induced modifications of the atomic velocity distribution. This results in line shape asymmetry, which can be characterized by incorporating a dispersive function into the Lorentzian profile \cite{Grimm1989}. The asymmetry was minimized by filling the cell with $\SI{1}{\milli\bar}$ argon buffer gas to restore the Maxwell Boltzmann distribution by 'strong' collisions. In 'strong' collisions, the change of velocity is approximately on the order of $v_0$, where $v_0$ represents the velocity of the class of atoms interacting with the pump beam. In precision measurements, it is essential to exercise caution in selecting the type of buffer gas, as a too high number of 'weak' collisions may cause a Gaussian lineshape of the Lamb dip \cite{Xu1983}. By ensuring equal intensity between the probe and pump beams, we can effectively avoid this asymmetry.

Ca-Ar collisions cause a line broadening $\gamma_\mathrm{c}$ and a shift $\delta_\mathrm{c}$ \cite{Hindmarsh1973,Farr1971}. While the broadening insignificantly deteriorates the line center determination, the shift has to be corrected for. It is calculated as 
\begin{equation}
\delta_\mathrm{c} = \sigma_\mathrm{Ca-Ar} \cdot \overline{v} \cdot n_\mathrm{Ar} \, ,    
\end{equation}
where $\sigma_\mathrm{Ca-Ar}$ denotes the collisional cross section, which is precisely measured in Ref.~\cite{Smith1972}, $\overline{v}$ the mean relative collision velocity, and $n_\mathrm{Ar}$ the argon density. For each Ca isotope, we calculate 
\begin{equation}
\overline{v} = \sqrt{8k_\mathrm{B}T/(3\pi m_\mathrm{red})}    
\end{equation}
with
\begin{equation}
m_\mathrm{red} = M_\mathrm{A} m_\mathrm{Ar}\,/\,(M_\mathrm{A}+m_\mathrm{Ar})    
\end{equation}
to accommodate the specific velocity distribution of each isotope $A$ with its mass $M_\mathrm{A}$ \cite{Hindmarsh1973,Farr1971}.

Recoil effects are not measurable, as the atomic linewidth is much larger than the recoil splitting of the line. With Ref.~\cite{Hall1976}, it can be shown that the recoil splitting is symmetrical around the resonance up to $\SI{800}{\hertz}$. We expect this to be common mode for all resonances, but nevertheless an error of $\SI{1}{\kilo\hertz}$ is added to the uncertainty margin.

\section{Measurements}
\subsection{Modified Lorentzian lineshape}
It has been shown in previous theoretical and experimental work, that the Lamp dip profile is modified in the presence of collisions and thus exhibits asymmetry \cite{Walkup1980,Walkup1984,Ciuryto1997}. This originates from a finite collision time resulting in failure of the impact approximation, which requires $T_\mathrm{d} \cdot \Delta \ll 1\,$. Here, $T_\mathrm{d}$ denotes the finite collision time and $\Delta = \nu-\nu_0$ the laser detuning from the atomic resonance $\nu_0$ \cite{Ciuryto1997}. The actual lineshape that is used to evaluate the data from this experiment thus reads
\begin{equation}
I \left( \Delta \right) \propto  \frac{\gamma_\mathrm{N} + \gamma_\mathrm{c}\left(\Delta\right)}{\Delta^2 + \left[ \, \gamma_\mathrm{N} + \gamma_\mathrm{c}\left(0\right) \, \right]^2}
\label{eq:fit}
\end{equation}
\noindent
with

\begin{equation}
\gamma_\mathrm{c}\left(\Delta\right) = \gamma_\mathrm{c}\left(0\right) \cdot \left(  1 + a_1  \Delta T_\mathrm{d} \right)\, ,
\label{eq:gammac}
\end{equation}
whereas $\gamma_\mathrm{N}$ is the natural linewidth, $a_1$ is a potential-dependent parameter, and $\gamma_\mathrm{c}\left(0\right)$ the collisional broadening from the impact approximation \cite{Walkup1980,Walkup1984}. To fit the parameters, the background is approximated by a second-order polynomial, as the resonances of all isotopes lie on the wing of the $^{40}\mathrm{Ca}$ resonance. It turns out that the resonances of $^{43}\mathrm{Ca}$ and $^{44}\mathrm{Ca}$ contribute significantly to the background seen at the position of the $^{42}\mathrm{Ca}$ resonance. To address this, a comprehensive scan over the resonances of $^{42}\mathrm{Ca}$, $^{43}\mathrm{Ca}$, and $^{44}\mathrm{Ca}$ has been performed and is fitted in a heuristic way to determine a more precise background approximation.

\subsection{Lamb dip measurement}
In this work, the Lamb dips of $^{40}\mathrm{Ca}$, $^{42}\mathrm{Ca}$, $^{44}\mathrm{Ca}$, and $^{48}\mathrm{Ca}$ are measured. Their abundances are $\SI{97.0}{\percent}$, $\SI{0.6}{\percent}$, $\SI{2.1}{\percent}$ and $\SI{0.2}{\percent}$, respectively. The hyperfine components of $^{43}\mathrm{Ca}$ are not determined, as the lines are not well separated from the crossover resonances. The abundance of $^{46}\mathrm{Ca}$ is $4\cdot\SI{e-3}{\percent}$, which is too low to determine its line center with competitive precision \cite{Kondev2021}.

Each isotope undergoes multiple scans in opposite scan directions. For each scan, Eq.~\ref{eq:fit} is fitted more than 100 times to the data, each time with a different fit range. The extracted line centers are averaged and the standard deviation of their distribution is considered as the statistical error of this measurement. 

The experimental stability is confirmed, as the extracted line centers for each scan are equal within their uncertainty interval. In between the measurements for $^{42}\mathrm{Ca}$, $^{44}\mathrm{Ca}$, and $^{48}\mathrm{Ca}$, the resonance frequency of $^{40}\mathrm{Ca}$ is determined to ensure that the measurement precision does not suffer from possible drifts of the ULE cavity resonance.

\begin{figure*}
\includegraphics[width=\textwidth]{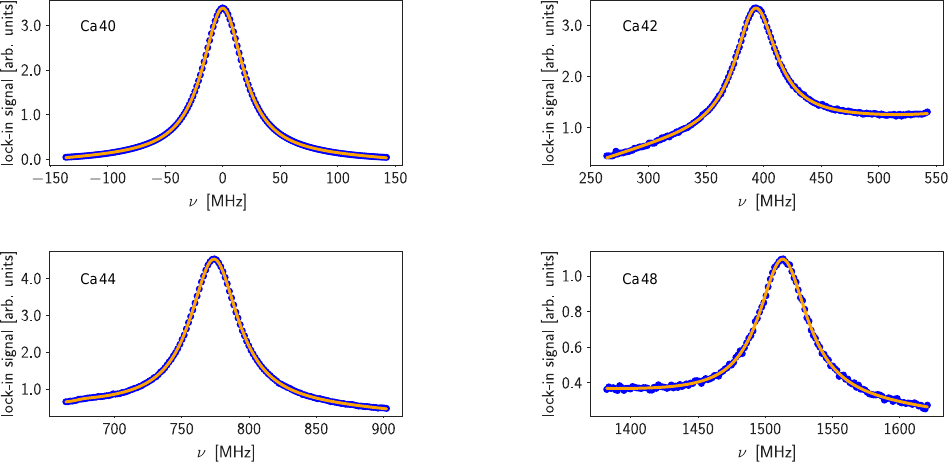}
\caption{The figure shows the demodulated and low pass filtered lock-in signal exhibiting the measured Lamb dips of $^{40}\mathrm{Ca}$, $^{42}\mathrm{Ca}$, $^{44}\mathrm{Ca}$, and $^{48}\mathrm{Ca}$. $\,\nu$ denotes the frequency offset from $^{40}\mathrm{Ca}$ in natural frequencies, while the lock-in signal is depicted in arbitrary units. The resonances are scaled by a factor of $100$, $50$, and $300$ for $^{42}\mathrm{Ca}$, $^{44}\mathrm{Ca}$, and $^{48}\mathrm{Ca}$, respectively, and the offset is adjusted for better visibility.}
\label{fig:res}
\end{figure*}

\begin{figure}[h]
\includegraphics[width=0.42\textwidth]{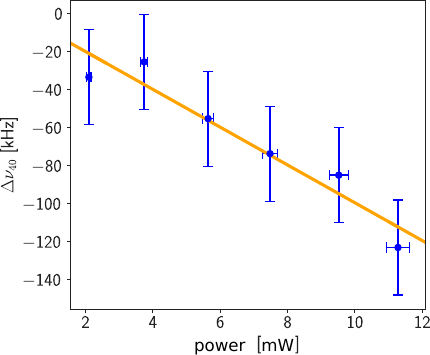}
\caption{The resonance of $^{40}\mathrm{Ca}$ is measured for different light powers and is shown as deviation from the extrapolated linecenter at zero intensity. An error of $\SI{2}{\percent}$ is estimated for the power measurement, while the displayed uncertainty of $\Delta\nu_{40}$ originates from the fit. The slope of the fitted linear function serves as AC Stark shift parameter.}
\label{fig:acstark}
\end{figure}

To correct the measured line centers for AC Stark shifts, the resonance frequency of $^{40}\mathrm{Ca}$ is measured for different light powers. The data is shown in Fig.~\ref{fig:acstark}. By fitting a linear function, the slope is determined to be $-10.0(1.5)\,\SI{}{\kilo\hertz\per(\milli\watt\per\centi\meter^2)}$. The corrections on the isotope shifts are smaller than $\SI{6}{\kilo\hertz}$ and thus are one order of magnitude lower than the statistical errors. Compared to the AC Stark shift parameter of Ref.~\cite{Salumbides2011}, our value is approximately a factor of $5$ lower and has the opposite sign. This indicates that other effects related to power significantly disturb the determination of the shift parameter. To improve the precision of our measurements, a more precise work on power dependent shifts would be necessary to rule out all possible sources of shifts, which goes beyond the scope of this work. 

Our measured isotope shifts are presented in Tab.~\ref{tab:shifts}. The results agree well with Ref.~\cite{Salumbides2011}, but the uncertainty ranges are smaller by a factor of about $5$. The corresponding corrections and uncertainty budgets are displayed in table \ref{tab:uncertainty}. The Lamb dips are shown for each isotopes in Fig.~\ref{fig:res}.

\begin{table}[h]
\caption{\label{tab:shifts}%
Measured isotopes shifts relative to $^{40}\mathrm{Ca}$. and comparison to previously reported values. All values are in MHz. 
}
\begin{ruledtabular}
\begin{tabular}{cddd}
\textrm{Isotope}&
\multicolumn{1}{c}{\textrm{This work}}&
\multicolumn{1}{c}{\textrm{\cite{Salumbides2011}}}&
\multicolumn{1}{c}{\textrm{\cite{Nörtershäuser1998}}}\\
\midrule
$^{42}\mathrm{Ca}$ & 393.090(88) & 393.1(4) & 393.50(10)\\
$^{44}\mathrm{Ca}$ & 773.814(48) & 773.8(2) & 773.80(15)\\
$^{48}\mathrm{Ca}$ & 1512.989(73) & 1513.1(4) & 1513.00(20)\\
\end{tabular}
\end{ruledtabular}
\end{table}

\begin{table}[h]
\caption{\label{tab:uncertainty}%
Corrections and uncertainties of the isotope shift measurement. All values are in $\SI{}{\kilo\hertz}.$
}
\begin{ruledtabular}
\begin{tabular}{ldd}
\textrm{Source}&
\multicolumn{1}{c}{\textrm{Correction}}&
\multicolumn{1}{c}{\textrm{Uncertainty}}\\
\midrule
residual Doppler & 0 & 10 \\
second-order Doppler & 0 & <0.5 \\
recoil &   0 & 1 \\
AC Stark &  <5 & <1 \\
Zeeman effect &  0 & 0 \\
collisions &  <400 & <10 \\
cavity drift & 0 & 0 \\
statistical error &  0 & <85 \\

\end{tabular}
\end{ruledtabular}
\end{table}

\subsection{King plot analysis}

\begin{figure}
\includegraphics[width=0.49\textwidth]{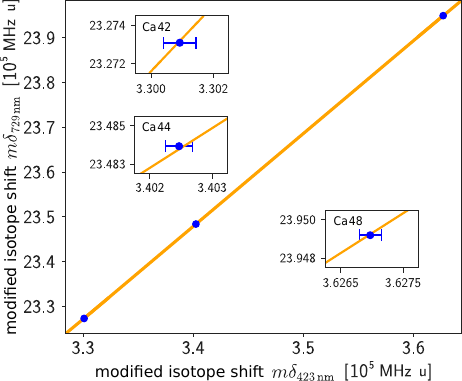}
\caption{King plot generated with our measurements and isotopes shifts of the $4s\, ^2S_{1/2} \rightarrow 3d\, ^2D_{5/2}$ transition at $\SI{729}{\nano\meter}$ in trapped $\mathrm{Ca}^+$ \cite{Knollmann2019}. The errorbars of $\delta m_{729\,\mathrm{nm}}$ are too small to be visible.}
\label{fig:kingplot}
\end{figure}

To further investigate our data, a King plot analysis is performed, as shown in Fig.~\ref{fig:kingplot}. We use isotope shifts of the $4s\, ^2S_{1/2} \rightarrow 3d\, ^2D_{5/2}$ quadrupole transition at $\SI{729}{\nano\meter}$ in trapped $\mathrm{Ca}^+$ with uncertainties below $\SI{10}{Hz}$ \cite{Knollmann2019}. 

In our King plots, modified isotope shifts $m\delta_\mathrm{729\,\mathrm{nm}}$ are plotted against our measurement $m\delta_\mathrm{423\,\mathrm{nm}}$, with the actual shift being scaled by $g^{A,A'} = \left( 1/m_\mathrm{A'} -1/m_\mathrm{A} \right)^{-1}$. Here, $m_\mathrm{A}$ and $m_\mathrm{A'}$ denote the nuclear masses of the respective isotopes $\mathrm{A}$ and $\mathrm{A'}$ \cite{Berengut2018}. For each isotope, the masses are calculated from the respective atomic mass \cite{Wang2017} as
\begin{equation}
    m_\mathrm{nucleus} = m_\mathrm{atom} - 20\, m_\mathrm{e} + \sum_{i=1}^{20}E_i^\mathrm{bind} \, .
\end{equation}
Here, $m_\mathrm{e}$ denotes the electron mass and $E_\mathrm{i}^\mathrm{bind}$ the binding energy of each electron $i$, which is derived from the NIST database \cite{Solaro2020, NIST_ASD}.

A linear function
\begin{equation}
    m\delta_\mathrm{729\,\mathrm{nm}} = K_{729\,\mathrm{nm}} - \frac{F_{729\,\mathrm{nm}}}{F_{423\,\mathrm{nm}}} K_{423\,\mathrm{nm}} +  \frac{F_{729\,\mathrm{nm}}}{F_{423\,\mathrm{nm}}}\,  m\delta_{423\,\mathrm{nm}}\,
\end{equation}
is fitted to the data, with $F_{i}$ and $K_{i}$ the field and mass shift parameters of the transition $i$, respectively \cite{Berengut2018}.

Our King plot reveals an excellent linearity between our measurement and Ref.~\cite{Knollmann2019}, as the line intersects all uncertainty intervals of our data points. From the fit, we find $F_{729\,\mathrm{nm}}/F_{423\,\mathrm{nm}} = \SI{2.07336(0.00085)}{}$ and $K_{729\,\mathrm{nm}} - ({F_{729\,\mathrm{nm}}}/{F_{423\,\mathrm{nm}}}) K_{423\,\mathrm{nm}} = 16.4290(29)\cdot\SI{e5}{\mega\hertz \atomicmassunit}\,$.

From isotope shift measurements of the $^2S_{1/2}\rightarrow {}^{2}P_{1/2}$ and $^2D_{3/2}\rightarrow {}^{2}P_{1/2}$ transitions in trapped $\mathrm{Ca}^+$, Ref.~\cite{Gebert2015} determines the field and mass shift coefficients for the respective transitions. Using their data to perform a King plot analysis, we can deduce the field and mass shift coefficients $F_{423\,\mathrm{nm}} = -178.4(3.2) \,\SI{}{\mega\hertz\per\femto\meter^2}$ and $K_{423\,\mathrm{nm}} = 363.6(1.4)\,\SI{}{\giga\hertz\atomicmassunit}$.\\ \\

\section{Conclusions and Outlook}

We have reported isotope shifts for the most abundant isotopes, $^{42}\mathrm{Ca}$, $^{44}\mathrm{Ca}$, and $^{48}\mathrm{Ca}$, relative to $^{40}\mathrm{Ca}$. Our measurement agrees very well with Ref.~\cite{Salumbides2011} and reduces the uncertainty by almost a factor of five to the permille scale of the linewidth.

By performing a King plot analysis with data from Ref.~\cite{Knollmann2019}, we find excellent linearity. Furthermore, by comparison to the field- and mass shift parameters determined in \cite{Gebert2015}, the field- and mass shift constants are calculated for the $423$-$\SI{}{\nano\meter}$ line.

In future measurements, the SNR can be improved by several orders of magnitude. As demonstrated in Ref.~\cite{Bjorklund2021}, double modulation schemes with an amplitude-modulated pump beam in modulation transfer spectroscopy can be applied. This is advantageous, as the frequency can be modulated with several $\SI{100}{\mega\hertz}$. Typical residual amplitude modulation noise from the frequency modulation occurs at the modulation frequency itself and thus can be mitigated by double modulation. With successful implementation, uncertainties of approximately $\SI{50}{\kilo\hertz}$ might be possible for all even-mass isotopes, including $^{46}\mathrm{Ca}$.

Further, changing to copropagating pump and probe beams eliminates the crossover resonances and allows for precise determination of the hyperfine coupling constants in $^{43}\mathrm{Ca}$ \cite{Banerjee2003}.

\section{Acknowledgments}
We acknowledge fruitful discussions with W.~Alt and F.~Vewinger and thank F.~Wenger and A.~Hamer for assistance in the setup and operation of the experiment. Financial Support from DFG through the Cluster of Excellence ML4Q (EXC2004/1 – 390534769) and projects 496941189 and INST 217/978-1 FUGG, as well as from BMBF (project 13N16371 QuantumGuide), is gratefully acknowledged. Further, H.~K.~and S.~St.~acknowledge funding by the state of North Rhine-Westphalia through the EIN Quantum NRW program.

Experimental data will be made available upon reasonable request.

\bibliography{Ca_423nm_spectroscopy}

\begin{thebibliography}{45}%
\makeatletter
\providecommand \@ifxundefined [1]{%
 \@ifx{#1\undefined}
}%
\providecommand \@ifnum [1]{%
 \ifnum #1\expandafter \@firstoftwo
 \else \expandafter \@secondoftwo
 \fi
}%
\providecommand \@ifx [1]{%
 \ifx #1\expandafter \@firstoftwo
 \else \expandafter \@secondoftwo
 \fi
}%
\providecommand \natexlab [1]{#1}%
\providecommand \enquote  [1]{``#1''}%
\providecommand \bibnamefont  [1]{#1}%
\providecommand \bibfnamefont [1]{#1}%
\providecommand \citenamefont [1]{#1}%
\providecommand \href@noop [0]{\@secondoftwo}%
\providecommand \href [0]{\begingroup \@sanitize@url \@href}%
\providecommand \@href[1]{\@@startlink{#1}\@@href}%
\providecommand \@@href[1]{\endgroup#1\@@endlink}%
\providecommand \@sanitize@url [0]{\catcode `\\12\catcode `\$12\catcode `\&12\catcode `\#12\catcode `\^12\catcode `\_12\catcode `\%12\relax}%
\providecommand \@@startlink[1]{}%
\providecommand \@@endlink[0]{}%
\providecommand \url  [0]{\begingroup\@sanitize@url \@url }%
\providecommand \@url [1]{\endgroup\@href {#1}{\urlprefix }}%
\providecommand \urlprefix  [0]{URL }%
\providecommand \Eprint [0]{\href }%
\providecommand \doibase [0]{https://doi.org/}%
\providecommand \selectlanguage [0]{\@gobble}%
\providecommand \bibinfo  [0]{\@secondoftwo}%
\providecommand \bibfield  [0]{\@secondoftwo}%
\providecommand \translation [1]{[#1]}%
\providecommand \BibitemOpen [0]{}%
\providecommand \bibitemStop [0]{}%
\providecommand \bibitemNoStop [0]{.\EOS\space}%
\providecommand \EOS [0]{\spacefactor3000\relax}%
\providecommand \BibitemShut  [1]{\csname bibitem#1\endcsname}%
\let\auto@bib@innerbib\@empty
\bibitem [{\citenamefont {Demtr{\"o}der}(2015)}]{demtröder2015laser2}%
  \BibitemOpen
  \bibfield  {author} {\bibinfo {author} {\bibfnamefont {W.}~\bibnamefont {Demtr{\"o}der}},\ }\href {https://books.google.de/books?id=13gaBgAAQBAJ} {\emph {\bibinfo {title} {Laser Spectroscopy 2: Experimental Techniques}}}\ (\bibinfo  {publisher} {Springer Berlin Heidelberg},\ \bibinfo {year} {2015})\BibitemShut {NoStop}%
\bibitem [{\citenamefont {Ludlow}\ \emph {et~al.}(2015)\citenamefont {Ludlow}, \citenamefont {Boyd}, \citenamefont {Ye}, \citenamefont {Peik},\ and\ \citenamefont {Schmidt}}]{Ludlow2015}%
  \BibitemOpen
  \bibfield  {author} {\bibinfo {author} {\bibfnamefont {A.~D.}\ \bibnamefont {Ludlow}}, \bibinfo {author} {\bibfnamefont {M.~M.}\ \bibnamefont {Boyd}}, \bibinfo {author} {\bibfnamefont {J.}~\bibnamefont {Ye}}, \bibinfo {author} {\bibfnamefont {E.}~\bibnamefont {Peik}},\ and\ \bibinfo {author} {\bibfnamefont {P.~O.}\ \bibnamefont {Schmidt}},\ }\bibfield  {title} {\bibinfo {title} {Optical atomic clocks},\ }\href {https://doi.org/10.1103/RevModPhys.87.637} {\bibfield  {journal} {\bibinfo  {journal} {Rev. Mod. Phys.}\ }\textbf {\bibinfo {volume} {87}},\ \bibinfo {pages} {637} (\bibinfo {year} {2015})}\BibitemShut {NoStop}%
\bibitem [{\citenamefont {Knollmann}\ \emph {et~al.}(2019)\citenamefont {Knollmann}, \citenamefont {Patel},\ and\ \citenamefont {Doret}}]{Knollmann2019}%
  \BibitemOpen
  \bibfield  {author} {\bibinfo {author} {\bibfnamefont {F.~W.}\ \bibnamefont {Knollmann}}, \bibinfo {author} {\bibfnamefont {A.~N.}\ \bibnamefont {Patel}},\ and\ \bibinfo {author} {\bibfnamefont {S.~C.}\ \bibnamefont {Doret}},\ }\href {https://doi.org/10.1103/PhysRevA.100.022514} {\bibfield  {journal} {\bibinfo  {journal} {Phys. Rev. A}\ }\textbf {\bibinfo {volume} {100}},\ \bibinfo {pages} {022514} (\bibinfo {year} {2019})}\BibitemShut {NoStop}%
\bibitem [{\citenamefont {Micke}\ \emph {et~al.}(2020)\citenamefont {Micke}, \citenamefont {Leopold}, \citenamefont {King}, \citenamefont {Benkler}, \citenamefont {Spieß}, \citenamefont {Schmöger}, \citenamefont {Schwarz}, \citenamefont {López-Urrutia},\ and\ \citenamefont {Schmidt}}]{Micke2020}%
  \BibitemOpen
  \bibfield  {author} {\bibinfo {author} {\bibfnamefont {P.}~\bibnamefont {Micke}}, \bibinfo {author} {\bibfnamefont {T.}~\bibnamefont {Leopold}}, \bibinfo {author} {\bibfnamefont {S.~A.}\ \bibnamefont {King}}, \bibinfo {author} {\bibfnamefont {E.}~\bibnamefont {Benkler}}, \bibinfo {author} {\bibfnamefont {L.~J.}\ \bibnamefont {Spieß}}, \bibinfo {author} {\bibfnamefont {L.}~\bibnamefont {Schmöger}}, \bibinfo {author} {\bibfnamefont {M.}~\bibnamefont {Schwarz}}, \bibinfo {author} {\bibfnamefont {J.~R.~C.}\ \bibnamefont {López-Urrutia}},\ and\ \bibinfo {author} {\bibfnamefont {P.~O.}\ \bibnamefont {Schmidt}},\ }\href {https://doi.org/10.1038/s41586-020-1959-8} {\bibfield  {journal} {\bibinfo  {journal} {Nature}\ }\textbf {\bibinfo {volume} {578}},\ \bibinfo {pages} {60} (\bibinfo {year} {2020})}\BibitemShut {NoStop}%
\bibitem [{\citenamefont {Hur}\ \emph {et~al.}(2022)\citenamefont {Hur} \emph {et~al.}}]{Hur2022}%
  \BibitemOpen
  \bibfield  {author} {\bibinfo {author} {\bibfnamefont {J.}~\bibnamefont {Hur}} \emph {et~al.},\ }\href {https://link.aps.org/doi/10.1103/PhysRevLett.128.163201} {\bibfield  {journal} {\bibinfo  {journal} {Phys. Rev. Lett.}\ }\textbf {\bibinfo {volume} {128}},\ \bibinfo {pages} {163201} (\bibinfo {year} {2022})}\BibitemShut {NoStop}%
\bibitem [{\citenamefont {Ono}\ \emph {et~al.}(2022)\citenamefont {Ono}, \citenamefont {Saito}, \citenamefont {Ishiyama}, \citenamefont {Higomoto}, \citenamefont {Takano}, \citenamefont {Takasu}, \citenamefont {Yamamoto}, \citenamefont {Tanaka},\ and\ \citenamefont {Takahashi}}]{Ono2022}%
  \BibitemOpen
  \bibfield  {author} {\bibinfo {author} {\bibfnamefont {K.}~\bibnamefont {Ono}}, \bibinfo {author} {\bibfnamefont {Y.}~\bibnamefont {Saito}}, \bibinfo {author} {\bibfnamefont {T.}~\bibnamefont {Ishiyama}}, \bibinfo {author} {\bibfnamefont {T.}~\bibnamefont {Higomoto}}, \bibinfo {author} {\bibfnamefont {T.}~\bibnamefont {Takano}}, \bibinfo {author} {\bibfnamefont {Y.}~\bibnamefont {Takasu}}, \bibinfo {author} {\bibfnamefont {Y.}~\bibnamefont {Yamamoto}}, \bibinfo {author} {\bibfnamefont {M.}~\bibnamefont {Tanaka}},\ and\ \bibinfo {author} {\bibfnamefont {Y.}~\bibnamefont {Takahashi}},\ }\href {https://doi.org/10.1103/PhysRevX.12.021033} {\bibfield  {journal} {\bibinfo  {journal} {Phys. Rev. X}\ }\textbf {\bibinfo {volume} {12}},\ \bibinfo {pages} {021033} (\bibinfo {year} {2022})}\BibitemShut {NoStop}%
\bibitem [{\citenamefont {Chang}\ \emph {et~al.}(2024)\citenamefont {Chang}, \citenamefont {Awazi}, \citenamefont {Berengut}, \citenamefont {Fuchs},\ and\ \citenamefont {Doret}}]{Chang2024}%
  \BibitemOpen
  \bibfield  {author} {\bibinfo {author} {\bibfnamefont {T.~T.}\ \bibnamefont {Chang}}, \bibinfo {author} {\bibfnamefont {B.~B.}\ \bibnamefont {Awazi}}, \bibinfo {author} {\bibfnamefont {J.~C.}\ \bibnamefont {Berengut}}, \bibinfo {author} {\bibfnamefont {E.}~\bibnamefont {Fuchs}},\ and\ \bibinfo {author} {\bibfnamefont {S.~C.}\ \bibnamefont {Doret}},\ }\href@noop {} {} (\bibinfo {year} {2024}),\ \Eprint {https://arxiv.org/abs/2311.17337} {arXiv:2311.17337 [physics.atom-ph]} \BibitemShut {NoStop}%
\bibitem [{\citenamefont {Berengut}\ \emph {et~al.}(2011)\citenamefont {Berengut}, \citenamefont {Flambaum},\ and\ \citenamefont {Kava}}]{Berengut2011}%
  \BibitemOpen
  \bibfield  {author} {\bibinfo {author} {\bibfnamefont {J.~C.}\ \bibnamefont {Berengut}}, \bibinfo {author} {\bibfnamefont {V.~V.}\ \bibnamefont {Flambaum}},\ and\ \bibinfo {author} {\bibfnamefont {E.~M.}\ \bibnamefont {Kava}},\ }\href {https://doi.org/10.1103/PhysRevA.84.042510} {\bibfield  {journal} {\bibinfo  {journal} {Phys. Rev. A}\ }\textbf {\bibinfo {volume} {84}},\ \bibinfo {pages} {042510} (\bibinfo {year} {2011})}\BibitemShut {NoStop}%
\bibitem [{\citenamefont {Dreissen}\ \emph {et~al.}(2022)\citenamefont {Dreissen}, \citenamefont {Yeh}, \citenamefont {Fürst}, \citenamefont {Grensemann},\ and\ \citenamefont {Mehlstäubler}}]{Dreissen2022}%
  \BibitemOpen
  \bibfield  {author} {\bibinfo {author} {\bibfnamefont {L.~S.}\ \bibnamefont {Dreissen}}, \bibinfo {author} {\bibfnamefont {C.~H.}\ \bibnamefont {Yeh}}, \bibinfo {author} {\bibfnamefont {H.~A.}\ \bibnamefont {Fürst}}, \bibinfo {author} {\bibfnamefont {K.~C.}\ \bibnamefont {Grensemann}},\ and\ \bibinfo {author} {\bibfnamefont {T.~E.}\ \bibnamefont {Mehlstäubler}},\ }\href@noop {} {\bibfield  {journal} {\bibinfo  {journal} {Nat. Commun.}\ }\textbf {\bibinfo {volume} {13}},\ \bibinfo {pages} {7314} (\bibinfo {year} {2022})}\BibitemShut {NoStop}%
\bibitem [{\citenamefont {Ohayon}\ \emph {et~al.}(2022)\citenamefont {Ohayon} \emph {et~al.}}]{Ohayon2022}%
  \BibitemOpen
  \bibfield  {author} {\bibinfo {author} {\bibfnamefont {B.}~\bibnamefont {Ohayon}} \emph {et~al.},\ }\href {https://doi.org/10.1103/PhysRevLett.128.011802} {\bibfield  {journal} {\bibinfo  {journal} {Phys. Rev. Lett.}\ }\textbf {\bibinfo {volume} {128}},\ \bibinfo {pages} {011802} (\bibinfo {year} {2022})}\BibitemShut {NoStop}%
\bibitem [{\citenamefont {Delaunay}\ \emph {et~al.}(2017)\citenamefont {Delaunay}, \citenamefont {Ozeri}, \citenamefont {Perez},\ and\ \citenamefont {Soreq}}]{Delaunay2017}%
  \BibitemOpen
  \bibfield  {author} {\bibinfo {author} {\bibfnamefont {C.}~\bibnamefont {Delaunay}}, \bibinfo {author} {\bibfnamefont {R.}~\bibnamefont {Ozeri}}, \bibinfo {author} {\bibfnamefont {G.}~\bibnamefont {Perez}},\ and\ \bibinfo {author} {\bibfnamefont {Y.}~\bibnamefont {Soreq}},\ }\href {https://doi.org/10.1103/PhysRevD.96.093001} {\bibfield  {journal} {\bibinfo  {journal} {Phys. Rev. D}\ }\textbf {\bibinfo {volume} {96}},\ \bibinfo {pages} {093001} (\bibinfo {year} {2017})}\BibitemShut {NoStop}%
\bibitem [{\citenamefont {Solaro}\ \emph {et~al.}(2020)\citenamefont {Solaro}, \citenamefont {Meyer}, \citenamefont {Fisher}, \citenamefont {Berengut}, \citenamefont {Fuchs},\ and\ \citenamefont {Drewsen}}]{Solaro2020}%
  \BibitemOpen
  \bibfield  {author} {\bibinfo {author} {\bibfnamefont {C.}~\bibnamefont {Solaro}}, \bibinfo {author} {\bibfnamefont {S.}~\bibnamefont {Meyer}}, \bibinfo {author} {\bibfnamefont {K.}~\bibnamefont {Fisher}}, \bibinfo {author} {\bibfnamefont {J.~C.}\ \bibnamefont {Berengut}}, \bibinfo {author} {\bibfnamefont {E.}~\bibnamefont {Fuchs}},\ and\ \bibinfo {author} {\bibfnamefont {M.}~\bibnamefont {Drewsen}},\ }\href {https://doi.org/10.1103/PhysRevLett.125.123003} {\bibfield  {journal} {\bibinfo  {journal} {Phys. Rev. Lett.}\ }\textbf {\bibinfo {volume} {125}},\ \bibinfo {pages} {123003} (\bibinfo {year} {2020})}\BibitemShut {NoStop}%
\bibitem [{\citenamefont {Counts}\ \emph {et~al.}(2020)\citenamefont {Counts} \emph {et~al.}}]{Counts2020}%
  \BibitemOpen
  \bibfield  {author} {\bibinfo {author} {\bibfnamefont {I.}~\bibnamefont {Counts}} \emph {et~al.},\ }\href {https://doi.org/10.1103/PhysRevLett.125.123002} {\bibfield  {journal} {\bibinfo  {journal} {Phys. Rev. Lett.}\ }\textbf {\bibinfo {volume} {125}},\ \bibinfo {pages} {123002} (\bibinfo {year} {2020})}\BibitemShut {NoStop}%
\bibitem [{\citenamefont {Berengut}\ \emph {et~al.}(2020)\citenamefont {Berengut}, \citenamefont {Delaunay}, \citenamefont {Geddes},\ and\ \citenamefont {Soreq}}]{Berengut2020}%
  \BibitemOpen
  \bibfield  {author} {\bibinfo {author} {\bibfnamefont {J.~C.}\ \bibnamefont {Berengut}}, \bibinfo {author} {\bibfnamefont {C.}~\bibnamefont {Delaunay}}, \bibinfo {author} {\bibfnamefont {A.}~\bibnamefont {Geddes}},\ and\ \bibinfo {author} {\bibfnamefont {Y.}~\bibnamefont {Soreq}},\ }\href {https://doi.org/10.1103/PhysRevResearch.2.043444} {\bibfield  {journal} {\bibinfo  {journal} {Phys. Rev. Res.}\ }\textbf {\bibinfo {volume} {2}},\ \bibinfo {pages} {043444} (\bibinfo {year} {2020})}\BibitemShut {NoStop}%
\bibitem [{\citenamefont {Rehbehn}\ \emph {et~al.}(2021)\citenamefont {Rehbehn} \emph {et~al.}}]{Rehbehn2021}%
  \BibitemOpen
  \bibfield  {author} {\bibinfo {author} {\bibfnamefont {N.-H.}\ \bibnamefont {Rehbehn}} \emph {et~al.},\ }\href {https://link.aps.org/doi/10.1103/PhysRevA.103.L040801} {\bibfield  {journal} {\bibinfo  {journal} {Phys. Rev. A}\ }\textbf {\bibinfo {volume} {103}} (\bibinfo {year} {2021})}\BibitemShut {NoStop}%
\bibitem [{\citenamefont {Figueroa}\ \emph {et~al.}(2022)\citenamefont {Figueroa}, \citenamefont {Berengut}, \citenamefont {Dzuba}, \citenamefont {Flambaum}, \citenamefont {Budker},\ and\ \citenamefont {Antypas}}]{Figueroa2022}%
  \BibitemOpen
  \bibfield  {author} {\bibinfo {author} {\bibfnamefont {N.~L.}\ \bibnamefont {Figueroa}}, \bibinfo {author} {\bibfnamefont {J.~C.}\ \bibnamefont {Berengut}}, \bibinfo {author} {\bibfnamefont {V.~A.}\ \bibnamefont {Dzuba}}, \bibinfo {author} {\bibfnamefont {V.~V.}\ \bibnamefont {Flambaum}}, \bibinfo {author} {\bibfnamefont {D.}~\bibnamefont {Budker}},\ and\ \bibinfo {author} {\bibfnamefont {D.}~\bibnamefont {Antypas}},\ }\href {https://doi.org/10.1103/PhysRevLett.128.073001} {\bibfield  {journal} {\bibinfo  {journal} {Phys. Rev. Lett.}\ }\textbf {\bibinfo {volume} {128}},\ \bibinfo {pages} {073001} (\bibinfo {year} {2022})}\BibitemShut {NoStop}%
\bibitem [{\citenamefont {Potvliege}\ \emph {et~al.}(2023)\citenamefont {Potvliege}, \citenamefont {Nicolson}, \citenamefont {Jones},\ and\ \citenamefont {Spannowsky}}]{Potvliege2023}%
  \BibitemOpen
  \bibfield  {author} {\bibinfo {author} {\bibfnamefont {R.~M.}\ \bibnamefont {Potvliege}}, \bibinfo {author} {\bibfnamefont {A.}~\bibnamefont {Nicolson}}, \bibinfo {author} {\bibfnamefont {M.~P.~A.}\ \bibnamefont {Jones}},\ and\ \bibinfo {author} {\bibfnamefont {M.}~\bibnamefont {Spannowsky}},\ }\href {https://doi.org/10.1103/PhysRevA.108.052825} {\bibfield  {journal} {\bibinfo  {journal} {Phys. Rev. A}\ }\textbf {\bibinfo {volume} {108}},\ \bibinfo {pages} {052825} (\bibinfo {year} {2023})}\BibitemShut {NoStop}%
\bibitem [{\citenamefont {Flambaum}\ \emph {et~al.}(2018)\citenamefont {Flambaum}, \citenamefont {Geddes},\ and\ \citenamefont {Viatkina}}]{Flambaum2017}%
  \BibitemOpen
  \bibfield  {author} {\bibinfo {author} {\bibfnamefont {V.~V.}\ \bibnamefont {Flambaum}}, \bibinfo {author} {\bibfnamefont {A.~J.}\ \bibnamefont {Geddes}},\ and\ \bibinfo {author} {\bibfnamefont {A.~V.}\ \bibnamefont {Viatkina}},\ }\href {https://doi.org/10.1103/PhysRevA.97.032510} {\bibfield  {journal} {\bibinfo  {journal} {Phys. Rev. A}\ }\textbf {\bibinfo {volume} {97}},\ \bibinfo {pages} {032510} (\bibinfo {year} {2018})}\BibitemShut {NoStop}%
\bibitem [{\citenamefont {King}(1963)}]{King1963}%
  \BibitemOpen
  \bibfield  {author} {\bibinfo {author} {\bibfnamefont {W.~H.}\ \bibnamefont {King}},\ }\href {https://doi.org/10.1364/JOSA.53.000638} {\bibfield  {journal} {\bibinfo  {journal} {J. Opt. Soc. Am.}\ }\textbf {\bibinfo {volume} {53}},\ \bibinfo {pages} {638} (\bibinfo {year} {1963})}\BibitemShut {NoStop}%
\bibitem [{\citenamefont {Berengut}\ \emph {et~al.}(2018)\citenamefont {Berengut} \emph {et~al.}}]{Berengut2018}%
  \BibitemOpen
  \bibfield  {author} {\bibinfo {author} {\bibfnamefont {J.~C.}\ \bibnamefont {Berengut}} \emph {et~al.},\ }\href {https://doi.org/10.1103/PhysRevLett.120.091801} {\bibfield  {journal} {\bibinfo  {journal} {Phys. Rev. Lett.}\ }\textbf {\bibinfo {volume} {120}},\ \bibinfo {pages} {091801} (\bibinfo {year} {2018})}\BibitemShut {NoStop}%
\bibitem [{\citenamefont {Door}\ \emph {et~al.}(2024)\citenamefont {Door} \emph {et~al.}}]{Door2024}%
  \BibitemOpen
  \bibfield  {author} {\bibinfo {author} {\bibfnamefont {M.}~\bibnamefont {Door}} \emph {et~al.},\ }\href@noop {} {} (\bibinfo {year} {2024}),\ \Eprint {https://arxiv.org/abs/2403.07792} {arXiv:2403.07792 [physics.atom-ph]} \BibitemShut {NoStop}%
\bibitem [{\citenamefont {Nakada}(2019)}]{Nakada2019}%
  \BibitemOpen
  \bibfield  {author} {\bibinfo {author} {\bibfnamefont {H.}~\bibnamefont {Nakada}},\ }\href {https://doi.org/10.1103/PhysRevC.100.044310} {\bibfield  {journal} {\bibinfo  {journal} {Phys. Rev. C}\ }\textbf {\bibinfo {volume} {100}},\ \bibinfo {pages} {044310} (\bibinfo {year} {2019})}\BibitemShut {NoStop}%
\bibitem [{\citenamefont {Mortensen}\ \emph {et~al.}(2004)\citenamefont {Mortensen}, \citenamefont {Lindballe}, \citenamefont {Jensen}, \citenamefont {Staanum}, \citenamefont {Voigt},\ and\ \citenamefont {Drewsen}}]{Mortensen2004}%
  \BibitemOpen
  \bibfield  {author} {\bibinfo {author} {\bibfnamefont {A.}~\bibnamefont {Mortensen}}, \bibinfo {author} {\bibfnamefont {J.~J.~T.}\ \bibnamefont {Lindballe}}, \bibinfo {author} {\bibfnamefont {I.~S.}\ \bibnamefont {Jensen}}, \bibinfo {author} {\bibfnamefont {P.}~\bibnamefont {Staanum}}, \bibinfo {author} {\bibfnamefont {D.}~\bibnamefont {Voigt}},\ and\ \bibinfo {author} {\bibfnamefont {M.}~\bibnamefont {Drewsen}},\ }\href {https://doi.org/10.1103/PhysRevA.69.042502} {\bibfield  {journal} {\bibinfo  {journal} {Phys. Rev. A}\ }\textbf {\bibinfo {volume} {69}},\ \bibinfo {pages} {042502} (\bibinfo {year} {2004})}\BibitemShut {NoStop}%
\bibitem [{\citenamefont {Nörtershäuser}\ \emph {et~al.}(1998)\citenamefont {Nörtershäuser}, \citenamefont {Trautmann}, \citenamefont {Wendt},\ and\ \citenamefont {Bushaw}}]{Nörtershäuser1998}%
  \BibitemOpen
  \bibfield  {author} {\bibinfo {author} {\bibfnamefont {W.}~\bibnamefont {Nörtershäuser}}, \bibinfo {author} {\bibfnamefont {N.}~\bibnamefont {Trautmann}}, \bibinfo {author} {\bibfnamefont {K.}~\bibnamefont {Wendt}},\ and\ \bibinfo {author} {\bibfnamefont {B.}~\bibnamefont {Bushaw}},\ }\href {https://doi.org/https://doi.org/10.1016/S0584-8547(98)00108-6} {\bibfield  {journal} {\bibinfo  {journal} {Spectrochimica Acta Part B: Atomic Spectroscopy}\ }\textbf {\bibinfo {volume} {53}},\ \bibinfo {pages} {709} (\bibinfo {year} {1998})}\BibitemShut {NoStop}%
\bibitem [{\citenamefont {Andl}\ \emph {et~al.}(1982)\citenamefont {Andl}, \citenamefont {Bekk}, \citenamefont {G\"oring}, \citenamefont {Hanser}, \citenamefont {Nowicki}, \citenamefont {Rebel}, \citenamefont {Schatz},\ and\ \citenamefont {Thompson}}]{Andl1982}%
  \BibitemOpen
  \bibfield  {author} {\bibinfo {author} {\bibfnamefont {A.}~\bibnamefont {Andl}}, \bibinfo {author} {\bibfnamefont {K.}~\bibnamefont {Bekk}}, \bibinfo {author} {\bibfnamefont {S.}~\bibnamefont {G\"oring}}, \bibinfo {author} {\bibfnamefont {A.}~\bibnamefont {Hanser}}, \bibinfo {author} {\bibfnamefont {G.}~\bibnamefont {Nowicki}}, \bibinfo {author} {\bibfnamefont {H.}~\bibnamefont {Rebel}}, \bibinfo {author} {\bibfnamefont {G.}~\bibnamefont {Schatz}},\ and\ \bibinfo {author} {\bibfnamefont {R.~C.}\ \bibnamefont {Thompson}},\ }\href {https://doi.org/10.1103/PhysRevC.26.2194} {\bibfield  {journal} {\bibinfo  {journal} {Phys. Rev. C}\ }\textbf {\bibinfo {volume} {26}},\ \bibinfo {pages} {2194} (\bibinfo {year} {1982})}\BibitemShut {NoStop}%
\bibitem [{\citenamefont {Salumbides}\ \emph {et~al.}(2011)\citenamefont {Salumbides}, \citenamefont {Maslinskas}, \citenamefont {Dildar}, \citenamefont {Wolf}, \citenamefont {van Duijn}, \citenamefont {Eikema},\ and\ \citenamefont {Ubachs}}]{Salumbides2011}%
  \BibitemOpen
  \bibfield  {author} {\bibinfo {author} {\bibfnamefont {E.~J.}\ \bibnamefont {Salumbides}}, \bibinfo {author} {\bibfnamefont {V.}~\bibnamefont {Maslinskas}}, \bibinfo {author} {\bibfnamefont {I.~M.}\ \bibnamefont {Dildar}}, \bibinfo {author} {\bibfnamefont {A.~L.}\ \bibnamefont {Wolf}}, \bibinfo {author} {\bibfnamefont {E.-J.}\ \bibnamefont {van Duijn}}, \bibinfo {author} {\bibfnamefont {K.~S.~E.}\ \bibnamefont {Eikema}},\ and\ \bibinfo {author} {\bibfnamefont {W.}~\bibnamefont {Ubachs}},\ }\href {https://doi.org/10.1103/PhysRevA.83.012502} {\bibfield  {journal} {\bibinfo  {journal} {Phys. Rev. A}\ }\textbf {\bibinfo {volume} {83}},\ \bibinfo {pages} {012502} (\bibinfo {year} {2011})}\BibitemShut {NoStop}%
\bibitem [{\citenamefont {Gebert}\ \emph {et~al.}(2015)\citenamefont {Gebert}, \citenamefont {Wan}, \citenamefont {Wolf}, \citenamefont {Angstmann}, \citenamefont {Berengut},\ and\ \citenamefont {Schmidt}}]{Gebert2015}%
  \BibitemOpen
  \bibfield  {author} {\bibinfo {author} {\bibfnamefont {F.}~\bibnamefont {Gebert}}, \bibinfo {author} {\bibfnamefont {Y.}~\bibnamefont {Wan}}, \bibinfo {author} {\bibfnamefont {F.}~\bibnamefont {Wolf}}, \bibinfo {author} {\bibfnamefont {C.~N.}\ \bibnamefont {Angstmann}}, \bibinfo {author} {\bibfnamefont {J.~C.}\ \bibnamefont {Berengut}},\ and\ \bibinfo {author} {\bibfnamefont {P.~O.}\ \bibnamefont {Schmidt}},\ }\href {https://doi.org/10.1103/PhysRevLett.115.053003} {\bibfield  {journal} {\bibinfo  {journal} {Phys. Rev. Lett.}\ }\textbf {\bibinfo {volume} {115}},\ \bibinfo {pages} {053003} (\bibinfo {year} {2015})}\BibitemShut {NoStop}%
\bibitem [{\citenamefont {Kramida}(2020)}]{Kramida2020}%
  \BibitemOpen
  \bibfield  {author} {\bibinfo {author} {\bibfnamefont {A.}~\bibnamefont {Kramida}},\ }\href {https://doi.org/https://doi.org/10.1016/j.adt.2019.101322} {\bibfield  {journal} {\bibinfo  {journal} {Atomic Data and Nuclear Data Tables}\ }\textbf {\bibinfo {volume} {133-134}},\ \bibinfo {pages} {101322} (\bibinfo {year} {2020})}\BibitemShut {NoStop}%
\bibitem [{\citenamefont {Hänsch}\ and\ \citenamefont {Couillaud}(1980)}]{Hansch1980}%
  \BibitemOpen
  \bibfield  {author} {\bibinfo {author} {\bibfnamefont {T.}~\bibnamefont {Hänsch}}\ and\ \bibinfo {author} {\bibfnamefont {B.}~\bibnamefont {Couillaud}},\ }\href@noop {} {\bibfield  {journal} {\bibinfo  {journal} {Optics Communications}\ }\textbf {\bibinfo {volume} {35}},\ \bibinfo {pages} {441–444} (\bibinfo {year} {1980})}\BibitemShut {NoStop}%
\bibitem [{\citenamefont {Park}\ \emph {et~al.}(2001)\citenamefont {Park}, \citenamefont {Lee}, \citenamefont {Kwon},\ and\ \citenamefont {Cho}}]{Park2001}%
  \BibitemOpen
  \bibfield  {author} {\bibinfo {author} {\bibfnamefont {S.~E.}\ \bibnamefont {Park}}, \bibinfo {author} {\bibfnamefont {H.~S.}\ \bibnamefont {Lee}}, \bibinfo {author} {\bibfnamefont {T.~Y.}\ \bibnamefont {Kwon}},\ and\ \bibinfo {author} {\bibfnamefont {H.}~\bibnamefont {Cho}},\ }\href {https://doi.org/https://doi.org/10.1016/S0030-4018(01)01155-5} {\bibfield  {journal} {\bibinfo  {journal} {Optics Communications}\ }\textbf {\bibinfo {volume} {192}},\ \bibinfo {pages} {49} (\bibinfo {year} {2001})}\BibitemShut {NoStop}%
\bibitem [{\citenamefont {Alcock}\ \emph {et~al.}(1984)\citenamefont {Alcock}, \citenamefont {Itkin},\ and\ \citenamefont {Horrigan}}]{Alcock1984}%
  \BibitemOpen
  \bibfield  {author} {\bibinfo {author} {\bibfnamefont {C.~B.}\ \bibnamefont {Alcock}}, \bibinfo {author} {\bibfnamefont {V.~P.}\ \bibnamefont {Itkin}},\ and\ \bibinfo {author} {\bibfnamefont {M.~K.}\ \bibnamefont {Horrigan}},\ }\href {https://doi.org/10.1179/cmq.1984.23.3.309} {\bibfield  {journal} {\bibinfo  {journal} {Canadian Metallurgical Quarterly}\ }\textbf {\bibinfo {volume} {23}},\ \bibinfo {pages} {309} (\bibinfo {year} {1984})}\BibitemShut {NoStop}%
\bibitem [{\citenamefont {Grimm}\ and\ \citenamefont {Mlynek}(1989)}]{Grimm1989}%
  \BibitemOpen
  \bibfield  {author} {\bibinfo {author} {\bibfnamefont {R.}~\bibnamefont {Grimm}}\ and\ \bibinfo {author} {\bibfnamefont {J.}~\bibnamefont {Mlynek}},\ }\href {https://doi.org/10.1103/PhysRevLett.63.232} {\bibfield  {journal} {\bibinfo  {journal} {Phys. Rev. Lett.}\ }\textbf {\bibinfo {volume} {63}},\ \bibinfo {pages} {232} (\bibinfo {year} {1989})}\BibitemShut {NoStop}%
\bibitem [{\citenamefont {Tenenbaum}\ \emph {et~al.}(1983)\citenamefont {Tenenbaum}, \citenamefont {Miron}, \citenamefont {Lavi}, \citenamefont {Liran}, \citenamefont {Strauss}, \citenamefont {Oreg},\ and\ \citenamefont {Erez}}]{Xu1983}%
  \BibitemOpen
  \bibfield  {author} {\bibinfo {author} {\bibfnamefont {J.}~\bibnamefont {Tenenbaum}}, \bibinfo {author} {\bibfnamefont {E.}~\bibnamefont {Miron}}, \bibinfo {author} {\bibfnamefont {S.}~\bibnamefont {Lavi}}, \bibinfo {author} {\bibfnamefont {J.}~\bibnamefont {Liran}}, \bibinfo {author} {\bibfnamefont {M.}~\bibnamefont {Strauss}}, \bibinfo {author} {\bibfnamefont {J.}~\bibnamefont {Oreg}},\ and\ \bibinfo {author} {\bibfnamefont {G.}~\bibnamefont {Erez}},\ }\href {https://doi.org/10.1088/0022-3700/16/24/011} {\bibfield  {journal} {\bibinfo  {journal} {J. Phys. B: Atom. Mol. Phys.}\ }\textbf {\bibinfo {volume} {16}},\ \bibinfo {pages} {4543} (\bibinfo {year} {1983})}\BibitemShut {NoStop}%
\bibitem [{\citenamefont {Hindmarsh}\ and\ \citenamefont {Farr}(1973)}]{Hindmarsh1973}%
  \BibitemOpen
  \bibfield  {author} {\bibinfo {author} {\bibfnamefont {W.}~\bibnamefont {Hindmarsh}}\ and\ \bibinfo {author} {\bibfnamefont {J.~M.}\ \bibnamefont {Farr}},\ }\href {https://doi.org/https://doi.org/10.1016/0079-6727(73)90005-0} {\bibfield  {journal} {\bibinfo  {journal} {Progress in Quantum Electronics}\ }\textbf {\bibinfo {volume} {2}},\ \bibinfo {pages} {141} (\bibinfo {year} {1973})}\BibitemShut {NoStop}%
\bibitem [{\citenamefont {Farr}\ and\ \citenamefont {Hindmarsh}(1971)}]{Farr1971}%
  \BibitemOpen
  \bibfield  {author} {\bibinfo {author} {\bibfnamefont {J.~M.}\ \bibnamefont {Farr}}\ and\ \bibinfo {author} {\bibfnamefont {W.~R.}\ \bibnamefont {Hindmarsh}},\ }\href@noop {} {\bibfield  {journal} {\bibinfo  {journal} {J. Phys. B: Atom. Mol. Phys.}\ }\textbf {\bibinfo {volume} {4}},\ \bibinfo {pages} {568} (\bibinfo {year} {1971})}\BibitemShut {NoStop}%
\bibitem [{\citenamefont {Smith}(1972)}]{Smith1972}%
  \BibitemOpen
  \bibfield  {author} {\bibinfo {author} {\bibfnamefont {G.}~\bibnamefont {Smith}},\ }\href@noop {} {\bibfield  {journal} {\bibinfo  {journal} {J. Phys. B: Atom. Mol. Phys.}\ }\textbf {\bibinfo {volume} {5}} (\bibinfo {year} {1972})}\BibitemShut {NoStop}%
\bibitem [{\citenamefont {Hall}\ \emph {et~al.}(1976)\citenamefont {Hall}, \citenamefont {Bord\'e},\ and\ \citenamefont {Uehara}}]{Hall1976}%
  \BibitemOpen
  \bibfield  {author} {\bibinfo {author} {\bibfnamefont {J.~L.}\ \bibnamefont {Hall}}, \bibinfo {author} {\bibfnamefont {C.~J.}\ \bibnamefont {Bord\'e}},\ and\ \bibinfo {author} {\bibfnamefont {K.}~\bibnamefont {Uehara}},\ }\href {https://doi.org/10.1103/PhysRevLett.37.1339} {\bibfield  {journal} {\bibinfo  {journal} {Phys. Rev. Lett.}\ }\textbf {\bibinfo {volume} {37}},\ \bibinfo {pages} {1339} (\bibinfo {year} {1976})}\BibitemShut {NoStop}%
\bibitem [{\citenamefont {Walkup}\ \emph {et~al.}(1980)\citenamefont {Walkup}, \citenamefont {Spielfiedel},\ and\ \citenamefont {Pritchard}}]{Walkup1980}%
  \BibitemOpen
  \bibfield  {author} {\bibinfo {author} {\bibfnamefont {R.~E.}\ \bibnamefont {Walkup}}, \bibinfo {author} {\bibfnamefont {A.}~\bibnamefont {Spielfiedel}},\ and\ \bibinfo {author} {\bibfnamefont {D.~E.}\ \bibnamefont {Pritchard}},\ }\href {https://doi.org/10.1103/PhysRevLett.45.986} {\bibfield  {journal} {\bibinfo  {journal} {Phys. Rev. Lett.}\ }\textbf {\bibinfo {volume} {45}},\ \bibinfo {pages} {986} (\bibinfo {year} {1980})}\BibitemShut {NoStop}%
\bibitem [{\citenamefont {Walkup}\ \emph {et~al.}(1984)\citenamefont {Walkup}, \citenamefont {Stewart},\ and\ \citenamefont {Pritchard}}]{Walkup1984}%
  \BibitemOpen
  \bibfield  {author} {\bibinfo {author} {\bibfnamefont {R.}~\bibnamefont {Walkup}}, \bibinfo {author} {\bibfnamefont {B.}~\bibnamefont {Stewart}},\ and\ \bibinfo {author} {\bibfnamefont {D.~E.}\ \bibnamefont {Pritchard}},\ }\href {https://doi.org/10.1103/PhysRevA.29.169} {\bibfield  {journal} {\bibinfo  {journal} {Phys. Rev. A}\ }\textbf {\bibinfo {volume} {29}},\ \bibinfo {pages} {169} (\bibinfo {year} {1984})}\BibitemShut {NoStop}%
\bibitem [{\citenamefont {Ciurylo}\ \emph {et~al.}(1997)\citenamefont {Ciurylo}, \citenamefont {Szudy},\ and\ \citenamefont {Trawifiski}}]{Ciuryto1997}%
  \BibitemOpen
  \bibfield  {author} {\bibinfo {author} {\bibfnamefont {R.}~\bibnamefont {Ciurylo}}, \bibinfo {author} {\bibfnamefont {J.}~\bibnamefont {Szudy}},\ and\ \bibinfo {author} {\bibfnamefont {R.~S.}\ \bibnamefont {Trawifiski}},\ }\href@noop {} {\bibfield  {journal} {\bibinfo  {journal} {J. Quant. Spcctrosc. Radiat. Transfer}\ }\textbf {\bibinfo {volume} {57}},\ \bibinfo {pages} {551} (\bibinfo {year} {1997})}\BibitemShut {NoStop}%
\bibitem [{\citenamefont {Kondev}\ \emph {et~al.}(2021)\citenamefont {Kondev}, \citenamefont {Wang}, \citenamefont {Huang}, \citenamefont {Naimi},\ and\ \citenamefont {Audi}}]{Kondev2021}%
  \BibitemOpen
  \bibfield  {author} {\bibinfo {author} {\bibfnamefont {F.}~\bibnamefont {Kondev}}, \bibinfo {author} {\bibfnamefont {M.}~\bibnamefont {Wang}}, \bibinfo {author} {\bibfnamefont {W.}~\bibnamefont {Huang}}, \bibinfo {author} {\bibfnamefont {S.}~\bibnamefont {Naimi}},\ and\ \bibinfo {author} {\bibfnamefont {G.}~\bibnamefont {Audi}},\ }\href {https://doi.org/10.1088/1674-1137/abddae} {\bibfield  {journal} {\bibinfo  {journal} {Chinese Physics C}\ }\textbf {\bibinfo {volume} {45}},\ \bibinfo {pages} {030001} (\bibinfo {year} {2021})}\BibitemShut {NoStop}%
\bibitem [{\citenamefont {Wang}\ \emph {et~al.}(2017)\citenamefont {Wang}, \citenamefont {G.}, \citenamefont {Kondev}, \citenamefont {W.J.}, \citenamefont {Naimi},\ and\ \citenamefont {Xing}}]{Wang2017}%
  \BibitemOpen
  \bibfield  {author} {\bibinfo {author} {\bibfnamefont {M.}~\bibnamefont {Wang}}, \bibinfo {author} {\bibfnamefont {A.}~\bibnamefont {G.}}, \bibinfo {author} {\bibfnamefont {F.~G.}\ \bibnamefont {Kondev}}, \bibinfo {author} {\bibfnamefont {H.}~\bibnamefont {W.J.}}, \bibinfo {author} {\bibfnamefont {S.}~\bibnamefont {Naimi}},\ and\ \bibinfo {author} {\bibfnamefont {X.}~\bibnamefont {Xing}},\ }\href {https://doi.org/10.1088/1674-1137/41/3/030003} {\bibfield  {journal} {\bibinfo  {journal} {Chinese Physics C}\ }\textbf {\bibinfo {volume} {41}},\ \bibinfo {pages} {030003} (\bibinfo {year} {2017})}\BibitemShut {NoStop}%
\bibitem [{\citenamefont {Kramida}\ \emph {et~al.}(2023)\citenamefont {Kramida}, \citenamefont {{Yu.~Ralchenko}}, \citenamefont {Reader},\ and\ \citenamefont {{and NIST ASD Team}}}]{NIST_ASD}%
  \BibitemOpen
  \bibfield  {author} {\bibinfo {author} {\bibfnamefont {A.}~\bibnamefont {Kramida}}, \bibinfo {author} {\bibnamefont {{Yu.~Ralchenko}}}, \bibinfo {author} {\bibfnamefont {J.}~\bibnamefont {Reader}},\ and\ \bibinfo {author} {\bibnamefont {{and NIST ASD Team}}},\ }\href@noop {} {}\bibinfo {howpublished} {NIST Atomic Spectra Database (ver. 5.11),} (\bibinfo {year} {2023})\BibitemShut {NoStop}%
\bibitem [{\citenamefont {Bjorklund}\ and\ \citenamefont {Whittaker}(2021)}]{Bjorklund2021}%
  \BibitemOpen
  \bibfield  {author} {\bibinfo {author} {\bibfnamefont {G.~C.}\ \bibnamefont {Bjorklund}}\ and\ \bibinfo {author} {\bibfnamefont {E.~A.}\ \bibnamefont {Whittaker}},\ }\href {https://doi.org/10.1021/acs.jpca.1c05752} {\bibfield  {journal} {\bibinfo  {journal} {Journal of Physical Chemistry A}\ }\textbf {\bibinfo {volume} {125}},\ \bibinfo {pages} {8519} (\bibinfo {year} {2021})}\BibitemShut {NoStop}%
\bibitem [{\citenamefont {Banerjee}\ and\ \citenamefont {Natarajan}(2003)}]{Banerjee2003}%
  \BibitemOpen
  \bibfield  {author} {\bibinfo {author} {\bibfnamefont {A.}~\bibnamefont {Banerjee}}\ and\ \bibinfo {author} {\bibfnamefont {V.}~\bibnamefont {Natarajan}},\ }\href {https://opg.optica.org/ol/abstract.cfm?URI=ol-28-20-1912} {\bibfield  {journal} {\bibinfo  {journal} {Opt. Lett.}\ }\textbf {\bibinfo {volume} {28}},\ \bibinfo {pages} {1912} (\bibinfo {year} {2003})}\BibitemShut {NoStop}%
\end{thebibliography}%

\end{document}